\documentclass{iau}
\usepackage{graphicx}

\title[The magnetosphere of V4046 Sgr] 
{The magnetosphere of the close accreting PMS binary V4046 Sgr AB}

\author[S. G. Gregory et al.]   
{S. G. Gregory$^1$, V. R. Holzwarth$^2$, J.-F. Donati$^3$, G. A. J. Hussain$^4$, T. Montmerle$^5$,
E. Alecian$^6$, S. H. P. Alencar$^7$, C. Argiroffi$^8$, \\ 
M. Audard$^9$, J. Bouvier$^{10}$, F. Damiani$^8$,
M. G{\"u}del$^{11}$, \\ 
D. P. Huenemoerder$^{12}$, J. H. Kastner$^{13}$, A. Maggio$^8$, G. G. Sacco$^{14}$
\and G. A. Wade$^{15}$}

\affiliation{$^1$School of Physics \& Astronomy, University of St Andrews, St Andrews,
KY16 9SS, U.K. \\ email: {\tt sg64@st-andrews.ac.uk} \\[\affilskip]
$^2$Freytagstr. 7, D-79114 Freiburg i.Br., Germany \\[\affilskip] 
$^3$Inst. de Recherche en Astrophysique et Plan{\'e}tologie UMR 5277, Toulouse, FÐ31400 France \\[\affilskip]
$^4$ESO, Karl-Schwarzschild-Str. 2, D-85748 Garching, Germany \\[\affilskip]
$^5$Institut d'Astrophysique de Paris, 98bis bd Arago, FR-75014 Paris, France \\[\affilskip] 
$^6$Obs. de Paris, LESIA, 5, place Jules Janssen, F-92195 Meudon Principal Cedex, France \\[\affilskip] 
$^7$Dept. de F{\`i}sica - UFMG, Av. Ant{\^o}nio Carlos, 6627, 30270-901 Belo Horizonte, MG, Brazil \\[\affilskip] 
$^8$INAF-Osservatorio Astronomico di Palermo, Piazza del Parlamento 1, I-90134 Palermo, Italy \\[\affilskip] 
$^9$ISDC Data Center for Astrophysics, Univ. of Geneva, CH-1290 Versoix, Switzerland \\[\affilskip] 
$^{10}$UJF-Grenoble 1/CNRS-INSU, IPAG, UMR 5274, F-38041, Grenoble, France \\[\affilskip] 
$^{11}$Dept. of Astrophysics, University of Vienna, T{\"u}rkenschanzstrasse 17, A-1180 Vienna, Austria \\[\affilskip] 
$^{12}$MIT, Kavli Inst. for Astrophysics \& Space Research, Cambridge, MA 02139, U.S.A. \\[\affilskip] 
$^{13}$CIS, Rochester Inst. of Technology, 54 Lomb Memorial Drive, Rochester, NY 14623, U.S.A. \\[\affilskip] 
$^{14}$INAF-Arcetri Astrophysical Observatory, Largo Enrico Fermi 5, I - 50125 Florence, Italy \\[\affilskip] 
$^{15}$Dept. of Physics, Royal Military College of Canada, Kingston, K7K 7B4, Canada \\[\affilskip] 
}

\pubyear{2013}
\volume{302}  
\pagerange{xxx--xxx}
\jname{Magnetic fields throughout stellar evolution}
\editors{M. Jardine, P. Petit \& H. Spruit, eds.}
\begin{document}

\maketitle

\begin{abstract}
We present a preliminary 3D potential field extrapolation model of the joint magnetosphere of the close accreting
PMS binary V4046 Sgr.  The model is derived from magnetic maps obtained as
part of a coordinated optical and X-ray observing program.     
\end{abstract}

\firstsection 
\section{Introduction - large multi-wavelength observing campaign}
V4046 Sgr is a close (separation $\sim9\,{\rm R}_\odot$; \cite[Donati et al 2011]{don11}) circularised and synchronised PMS binary, accreting
gas from a large circumbinary disk \cite[(Rosenfeld et al. 2012)]{ros12}.  It was observed as part of a coordinated X-ray and spectropolarimetric observing program with {\it XMM-Newton} and ESPaDOnS@CFHT during 2009.  
The observational highlights include: (i). the derivation of the first magnetic maps of a close accreting PMS binary system, see Fig.\,\ref{maps} \cite[(Donati et al. 2011)]{don11}.  (ii). The detection of rotationally modulated soft X-ray emission associated with accretion shocks where accreting gas impacts the surface of the stars \cite[(Argiroffi et al. 2012)]{arg11}.  The modulation period is half of the binary orbital period.  (iii). The realisation that V4046 Sgr may
be a quadruple system, with GSC~07396-00759 a distant (projected separation $\sim$12,350$\,{\rm au}$) companion to V4046~Sgr~AB \cite[(Kastner et al. 2011)]{kas11}.  The companion itself is likely a non-accreting PMS binary. 

\section{The magnetic field of V4046 Sgr \& field extrapolation}
\begin{figure}
\begin{center}
 \includegraphics[width=72mm]{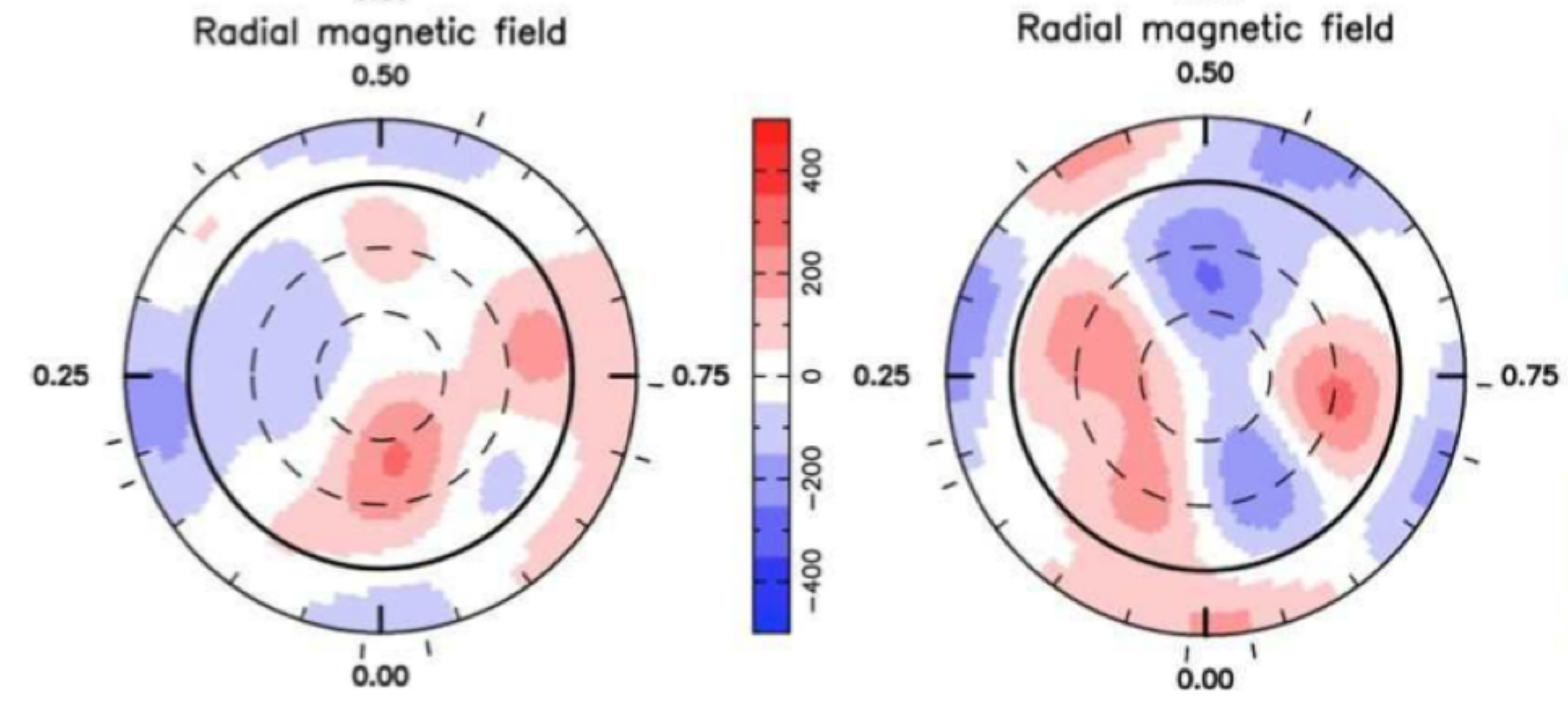} 
  \includegraphics[width=53mm]{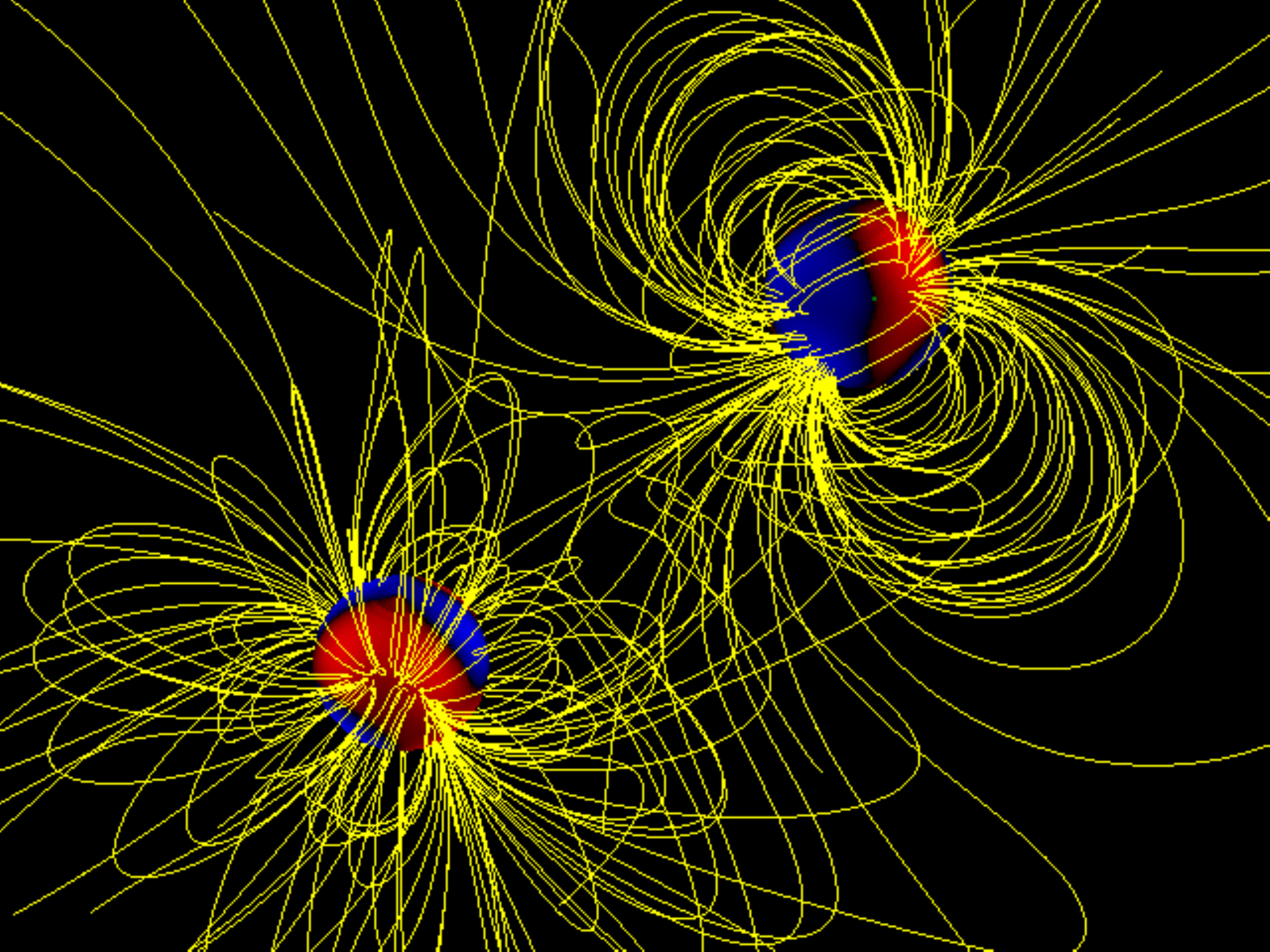} 
 \caption{Magnetic maps of the primary/secondary star of V4046 Sgr (left/middle respectively).  Blue/red is negative/postive field, with fluxes labelled in Gauss.  Tick marks/numbers denote the phases of observation/rotation phase.  The maps are shown in flattened polar
 projection with the bold circle/dashed lines the stellar equator/lines of constant latitude. Maps of the other field components \& brightness/excess accretion-related emission maps can be found in \cite[Donati et al. (2011)]{don11}.  The right panel shows a binary magnetic field extrapolation from the magnetic maps of V4046 Sgr. Only the large-scale field lines are shown. The
  magnetic fields are highly tilted with respect to the stellar rotation axes. Field lines connect through the interior of the binary from the nightside of
  one star to the dayside of the other.}
   \label{maps}
\end{center}
\end{figure}

Magnetic maps of V4046 Sgr, derived from 
Zeeman-Doppler imaging, are shown in Fig.\,\ref{maps} \cite[(Donati et al. 2011)]{don11}.  Only the radial field components are shown.  Both stars are 
found to host complex large-scale magnetic fields with weak dipole components, consistent with their partially convective internal 
structure \cite[(Gregory et al. 2012)]{gre12}.

A binary magnetic extrapolation is shown in Fig.\,\ref{maps}.  This has been constructed using a newly developed binary magnetic field extrapolation
code that will be described in a forthcoming paper (Holzwarth in prep.).  The code assumes that the large-scale field is potential and subject to three boundary conditions: the magnetic field is as measured from the maps at the surface of the each star, and a source surface boundary condition designed to mimic the pulling open of the large-scale magnetic loops by the stellar wind. The magnetic fields of both stars are linked, with loops connecting 
the dayside of one star to the nightside of the other.  The field geometry, and the distribution of accretion columns and hot spots, will be detailed in a future paper (Gregory et al. in prep.). 

{\it Acknowledgements}:
SGG acknowledges support from the Science \& Technology Facilities Council (STFC) via an Ernest 
Rutherford Fellowship [ST/J003255/1]. GAW is supported by a Discovery Grant from the Natural Science \& Engineering Research Council of Canada (NSERC).


\begin{thebibliography}{}
\bibitem[Argiroffi et al. (2012)]{arg12}{Argiroffi, C., Maggio, A., Montmerle, T., Huenemoerder, D.P., Alecian, E., Audard, M., Bouvier, J., Damiani, F., Donati, J.-F., Gregory, S.G., G{\"u}del, M., Hussain, G.A.J., Kastner, J.H., \& Sacco, G.G.} 2012, \textit{ApJ}, 752, 100
\bibitem[Donati et al. (2011)]{don11}
{Donati, J.-F., Gregory, S.G., Montmerle, T., Maggio, A., Argiroffi, C., Sacco, G., Hussain, G., Kastner, J., Alencar, S.H.P, Audard, M., Bouvier, J., 
Damiani, F., G{\"u}del, M., Huenemoerder, D., \& Wade, G.A.} 2011, \textit{MNRAS}, 417, 1747
\bibitem[Gregory et al. (2012)]{gre12}
{Gregory, S.G., Donati, J.-F., Morin, J., Hussain, G.A.J., Mayne, N.J., Hillenbrand, L.A., \& Jardine, M.} 2012, \textit{ApJ}, 755, 97
\bibitem[Kastner et al. (2011)]{kas11}
{Kastner, J.H., Sacco, G.G., Montez, R., Huenemoerder, D.P., Shi, H., Alecian, E., Argiroffi, C., Audard, M., 
	Bouvier, J., Damiani, F., Donati, J.-F., Gregory, S.G., 
	G{\"u}del, M., Hussain, G.A.J., Maggio, A., \& Montmerle, T.} 2011, \textit{ApJL}, 740, L17
\bibitem[Rosenfeld et al. (2012)]{ros12}{Rosenfeld, K.A., Andrews, S.M., Wilner, D.J. \& Stempels, H.C.} 2012, \textit{ApJ} 759, 119	
\end{thebibliography}
\end{document}